\documentclass[reprint,amsmath,amssymb,aps,
]{revtex4-2}

\usepackage{graphicx}
\usepackage{dcolumn}
\usepackage{bm}
\usepackage[utf8]{inputenc}
\usepackage{nicefrac}
\usepackage{natbib}
\usepackage{amsmath}
\usepackage[dvipsnames]{xcolor}
\usepackage[T1]{fontenc}
\usepackage{mathptmx}
\usepackage{etoolbox}
\newcommand{\RNum}[1]{\uppercase\expandafter{\romannumeral #1\relax}}
\makeatletter
\def\@email#1#2{%
 \endgroup
 \patchcmd{\titleblock@produce}
  {\frontmatter@RRAPformat}
  {\frontmatter@RRAPformat{\produce@RRAP{*#1\href{mailto:#2}{#2}}}\frontmatter@RRAPformat}
  {}{}
}%
\makeatother

\begin{document}
\preprint{APS/123-QED}

\title{Sympathetic cooling of trapped Th$^{3+}$ alpha-recoil ions for laser spectroscopy}

\author{G. Zitzer$^1$, J. Tiedau$^1$, M. V. Okhapkin$^1$, K. Zhang$^1$, C. Mokry$^{2,3}$, J. Runke$^{2,4}$, Ch. E. Düllmann$^{2,3,4}$, E. Peik$^{1}$}
 \affiliation{$^1$ Physikalisch-Technische Bundesanstalt, Braunschweig, Germany}
\affiliation{$^2$ Johannes Gutenberg University Mainz, Mainz, Germany}
\affiliation{$^3$ Helmholtz Institute Mainz, Mainz, Germany}
\affiliation{$^4$ GSI Helmholtzzentrum für Schwerionenforschung GmbH, Darmstadt, Germany}
\vspace{10pt}
\date{\today}

\begin{abstract}
Sympathetic cooling of Th$^{3+}$ ions is demonstrated in an experiment where $^{229}$Th and $^{230}$Th are extracted from uranium recoil ion sources and are confined in a linear Paul trap together with laser-cooled $^{88}$Sr$^+$ ions. 
Because of their similar charge-to-mass ratios the ions are closely coupled and arrange themselves in two-species Coulomb crystals, containing up to a few tens of Th$^{3+}$ ions. To show the suitability of the sympathetically cooled Th$^{3+}$ ions for high-resolution laser spectroscopy, the absolute frequencies and  isotope shifts of 5F$_{5/2}$\,$\rightarrow$\,6D$_{5/2}$ and 5F$_{7/2}$\,$\rightarrow$\,6D$_{5/2}$ transitions of $^{230}$Th$^{3+}$ have been measured. The system is developed for hyperfine spectroscopy of electronic transitions of nuclear ground and isomeric states in $^{229}$Th$^{3+}$.

\end{abstract}

\maketitle

\section{\label{sec:level1} Introduction}
There is a strong interest in laser spectroscopy of $^{229}$Th ions because of the low-energy (8.3\,eV) isomer ~\cite{Seiferle:2019,Kraemer:2022} that exists in this nucleus, at a transition energy that is more typical for resonances of valence electrons. This unique property of $^{229}$Th ions makes them promising candidates for experimental studies of largely unexplored links between atomic and nuclear physics like nuclear decay or excitation via electronic bridge processes or bound internal conversion ~\cite{Porsev:2010}. Being amenable to resonant laser excitation, this nuclear transition is attractive as a reference of an optical clock that combines high accuracy with a strong sensitivity for some hypothetical effects of new physics beyond the standard model that are now sought in frequency comparisons of atomic clocks ~\cite{Dmitriev:2009,Flambaum:2009,Peik:2021}. The thorium ion in the 3+ charge state is considered as the most promising candidate for such an optical nuclear clock based on trapped and laser-cooled ions ~\cite{Peik:2003,Campbell:2012,Beeks:2021}.

Trapping of $^{229}$Th ions in charge states 1+, 2+, and 3+ has been demonstrated with the ions produced in laser ablation from a solid target~\cite{Campbell:2011,Okhapkin:2015,Thielking:2018}, but the efficiency of this method decreases substantially with increasing charge~\cite{Zimmermann:2012}. Since $^{229}$Th is radioactive (half-life 7900 years) and scarce, it is desirable to work with an ion source requiring only a minimal amount of substance. While ion trapping experiments with $^{229}$Th$^{+}$ have been performed with a source containing only $10^{14}$ nuclei, corresponding to an activity below 1\,kBq \cite{Okhapkin:2015}, similar experiments with $^{229}$Th$^{3+}$ have only been reported with a source of approximately 100 kBq activity \cite{Campbell:2011}. An alternative and efficient approach to produce $^{229}$Th$^{3+}$ is the use of a thin layer of $^{233}$U that emits $^{229}$Th recoil ions in $\alpha$-decay~\cite{Wense:2016}. An additional advantage of the $^{229}$Th recoil ion source in comparison to laser ablation is that it also yields ions in the isomeric state $^{229m}$Th, albeit only in a small proportion of about 2\,\% \cite{Barci}. All studies of properties of the isomer to date have made use of sources based on radioactive decay. The thorium $\alpha$-recoil ions are emitted with an initial kinetic energy of up to 84\,keV and the most convenient method to dissipate this energy is by collisions with helium in a buffer gas cell, at room temperature at $10^3 - 10^4$~Pa pressure ~\cite{Neumar:2006}. In this process, the 3+ charge state of thorium is the highest one that is stable against charge exchange with He. In an ion trap coupled to such a recoil ion source via a de Laval nozzle, a Th$^{3+}$ storage time of about 100\,s has been observed, limited by reactions of the thorium ions with impurities in the buffer gas ~\cite{Wense:2016}. Longer storage times and the application of laser cooling require a more efficient separation of the ion trap from the buffer gas cell, for example by multiple stages of differential pumping or by a gate valve.

The level scheme of Th$^{3+}$ is suitable for laser cooling \cite{Peik:2003}, although the available near-infrared resonance transitions are relatively weak with calculated natural linewidths below 0.5~MHz ~\cite{Biemont:2004,Safronova:2006}. This limits the effectiveness of laser cooling in the initial stages after injection of the ions into the trap. In the case of $^{229}$Th$^{3+}$ with a nuclear spin $I=5/2$, laser cooling is additionally complicated by the complex hyperfine structure, making it necessary to apply several repumping wavelengths in order to avoid population trapping in sublevels that are not coupled to the lasers ~\cite{Campbell:2011}. 

The method of sympathetic cooling ~\cite{Larson:1986, Bowe:1999} of trapped ions uses a combination of an ion species that is amenable to laser cooling with an ion of interest that may not possess a suitable electronic transition for fast cyclic laser excitation.
Coupling between both species is provided by the Coulomb interaction and is particularly effective if a two-species Coulomb crystal can be formed. In this way the benefits of 
laser cooling like strong localization and reduced Doppler broadening become accessible for practically every ion species whose charge-to-mass ratio is not too different from one of the common coolant ions.
The method has already been used for ions of different elements and  charge states. For example, with thorium ions produced by laser ablation, sympathetic cooling of $^{229}$Th$^{3+}$ with the abundant isotope $^{232}$Th$^{3+}$ has been successfully applied ~\cite{Campbell:2011}, and $^{232}$Th$^{+}$ has been cooled with Ca$^+$ ~\cite{Groot-Berning:2019}. Sympathetic cooling of a doubly charged Ca$^{2+}$ ion with singly charged Ca$^+$ has been demonstrated~\cite{Kwapień:2007}. 
Coulomb crystallization of highly-charged $^{40}$Ar$^{13+}$ ions has been obtained by means of sympathetic cooling with laser-cooled Be$^{+}$ ions ~\cite{Schmöger:2015}. Extensive molecular-dynamics simulations of sympathetic cooling in multispecies ensembles composed of ions of different masses have been presented \cite{Zhang:2007}.

In this work we present a novel approach for precision spectroscopy of thorium ions, by realizing the trapping of $^{229}$Th$^{3+}$ and $^{230}$Th$^{3+}$ ($I=0$) $\alpha$-recoil ions in a linear radiofrequency trap in ultrahigh vacuum. The ions are cooled sympathetically by laser-cooled $^{88}$Sr$^+$ ions. This species has been selected as the coolant ion because of its convenient wavelengths and fast cooling transitions. We use the combination of singly charged and triply charged ions because their charge-to-mass ratios are similar, so that the two species are closely coupled within the same spatial region of the ion trap. As a result of the combination of buffer gas cooling and subsequent sympathetic laser cooling, the energy of the thorium ions is reduced by about eleven orders of magnitude from the recoil ion source to the ion trap.

\section{Experimental setup}
The experimental apparatus consists of an ion beamline for the production and guiding of thorium ions, a linear Paul trap, and an optical setup. The ion beamline is used for loading Th$^{3+}$ ions into the Paul trap. The optical setup consists of lasers for cooling of Sr$^{+}$ and for spectroscopy of Th$^{3+}$. The individual parts of the apparatus are described below.

\begin{figure}
\includegraphics[scale=0.42]{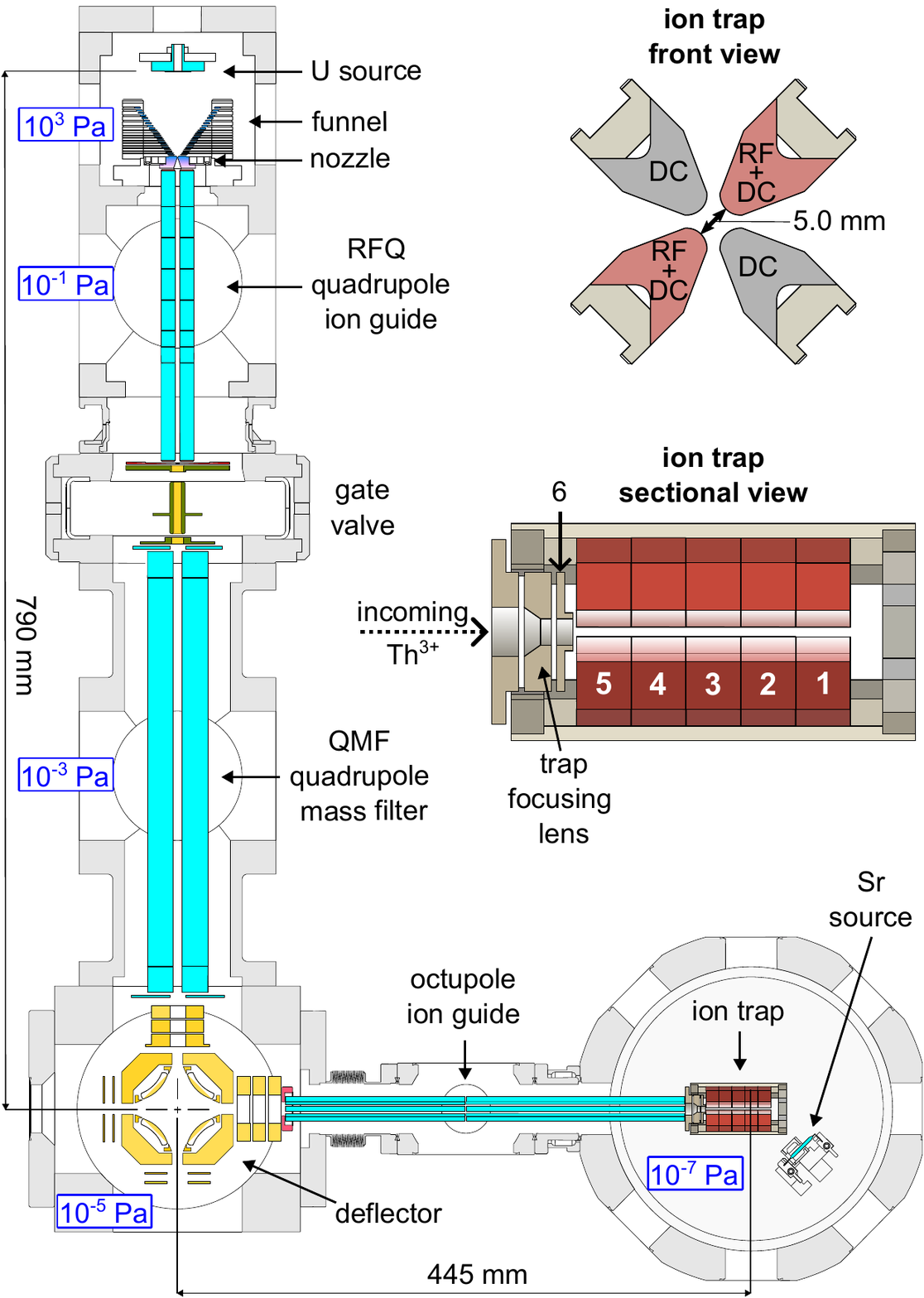}
\caption{\label{fig:UebersichtBeamline} Overview of the experimental system for loading and trapping of Th$^{3+}$ ions. Th$^{3+}$ ions are extracted from a uranium source.  In the subsequent RF quadrupole (RFQ) the ions are collected in bunches and passed to the quadrupole mass filter (QMF). The deflector redirects the ion beam 90° towards the ion trap. The octupole ion guide and the trap focusing lens are used to inject the ions into the trap. The trap segments are numbered from 1 to 5 starting from the end-cap section opposite to the octupole guide, and number 6 is the last part of the focusing lens. The sympathetic cooling of thorium with Sr$^{+}$ ions, loaded from a dispenser source, is performed in the second segment of the Paul trap. Helium pressures in the different sections indicate the performance of differential pumping when the gate valve is open.}
\end{figure}

\subsection{Th recoil ion beamline}
For loading the trap, the ions pass ion optics elements (see Fig.~\ref{fig:UebersichtBeamline}) which provide bunching, mass selection, and guiding.
As sources for $^{229}$Th and $^{230}$Th, two individual thin layers, one containing 10 kBq of $^{233}$U and one 10 kBq of $^{234}$U, were deposited on two silicon wafers over an area of 200~mm$^2$ each. A 100~nm-thick layer of Ti was applied to each wafer by sputtering, and the uranium layers were deposited on the Ti by the molecular plating method~\cite{Eberhardt:2018}. Thorium $\alpha$-recoil ions are emitted with energies up to 84\,keV and are decelerated in high-purity (research grade 6.0) helium buffer gas at about 3200\,Pa pressure. 
A liquid nitrogen trap and a heated getter element are used additionally to purify the buffer gas.
Guiding of the ions emitted from the U source is provided by a radiofrequency (RF) funnel~\cite{Shaffer:1997} mounted in front of the source. The RF electric field with a superimposed static field is applied to the convergent stacked 29 ring electrodes of the funnel. The amplitude of the RF voltage is $\sim$120\,V at a driving frequency of 780\,kHz.
The flow of the helium buffer gas in combination with the static electric field extracts the recoil ions from the source through a de Laval nozzle with a minimum diameter of 0.4\,mm and a voltage of about 3.5\,V into the next section. 
After passing the nozzle the ions are accumulated in a RF quadrupole (RFQ) consisting of eight segments. The RFQ has a total length of 226\,mm. The distance between opposite rods is 9.6\,mm with a rod diameter of 11\,mm. The RFQ driving voltage amplitude is $\sim$\,90\,V at 1.2\,MHz frequency. An additional longitudinal DC gradient of about 0.2\,V per RFQ segment is applied for an efficient accumulation of the ions. The ions in the RFQ are thermalized to room temperature by collisions with the helium buffer gas at $\approx$\,0.1\,Pa pressure. The RFQ vacuum section is the first stage of a differential pumping system towards the ion trap. 
The electrical configuration of the RFQ allows the extraction of ions in bunches or in a continuous beam. An aperture lens with an inner diameter of 2\,mm behind the last RFQ segment is used for efficient ion extraction. 

To separate the vacuum chamber of the ion trap from the flow of helium from the source chamber, an all-metal gate valve with an integrated einzel lens \cite{Harting:1976} in the orifice is mounted behind the RFQ. The einzel lens focuses the ions extracted from the RFQ into the downstream mass filter.
After closing the gate valve the pressure in the trap region decreases by an order of magnitude and reaches 10$^{-8}$\,Pa within a few seconds.
This reduces heating of the laser-cooled Sr$^+$ ions through collisions with background gas and increases the storage time of thorium ions. In addition, the valve simplifies maintenance of the beamline vacuum system. 

A quadrupole mass filter (QMF) \cite{Paul:1953} with Brubaker pre-filters \cite{Brubaker:1968} is used for the selection of the Th$^{3+}$ ions, filtering out other decay products from the uranium source, thorium ions in other charge states, and molecular ions formed in reactions with impurities in the buffer gas. The central segment of the QMF has a length of 300\,mm and 20 mm long Brubaker sections are attached on both sides of the central segment. The rod diameter of all electrodes is 20\,mm. The diagonal distance between two opposite electrodes is 17.7\,mm. Additional aperture lenses at the entrance and exit are used for focusing and extracting the ions. 
The QMF is typically operated with an RF voltage amplitude of $\approx$\,600\,V at $\approx$\,830\,kHz frequency with a DC voltage of about $-95$\,V. The measured resolution of the mass filter is \textit{m\,/\,$\Delta$m\,}$\approx$\,180 which allows efficient separation of $^{230}$Th$^{3+}$ from $^{234}$U$^{3+}$ and other daughter products of the uranium decay chain and molecular ions extracted from the RFQ. The second differential pumping stage improves the background pressure to $\approx$\,10$^{-3}$\,Pa in the QMF section. 

In order to provide laser access along the trap axis not obstructed by the ion source, the trajectories of the thorium ions are bent by 90$^\circ$ via an electrostatic deflector~\cite{Zeman}. Two pairs of electrodes, in combination with a set of shim electrodes, are applied with a DC voltage deflecting the incoming thorium ions from the mass filter toward the ion trap in a two-dimensional electrostatic quadrupole field. Individual sets of einzel lens assemblies are used for focusing the ion beam. When the deflector is not active, a channeltron detector that is mounted in the forward direction can be used for monitoring the ion flux.

An RF electric octupole guide is used to transport the ions from the deflector to the ion trap. Similar to the quadrupole ion guides, the electrodes of the octupole are connected to an RF ($\sim$\,1.2\,MHz) at about 110 V amplitude and to a common DC voltage of $-12$\,V. Use of the octupole configuration reduces transmission losses and simplifies the coupling to the ion guide \cite{Szabo:1986,Gerlich:1992}. 
The total length of the octupole is about 308\,mm. The minimum distance between two opposite electrodes is 15.3\,mm and the rod diameter is 4\,mm. An aperture and an einzel lens at the entrance and exit with a minimum inner diameter of up to 4.8\,mm are used to guide the ions into the trap.

\subsection{Ion trap}
For the experiments on laser cooling and spectroscopy, a linear radio-frequency trap (see Fig.~\ref{fig:UebersichtBeamline}) of 50\,mm total length is used. The trap consists of five segments with a length of 10\,mm each. The diagonal distance between opposite electrodes is 5\,mm and the radius of the electrodes is 2.9\,mm. 
Typically, a RF voltage in the range of 150 - 250\,V amplitude at a frequency of 2.8\,MHz is used, corresponding to a \textit{q} parameter in the range of $\textit{q}$ $\sim$ 0.2 - 0.3 (see e.g. Ref.~\cite{Paul:1990}).

The trap segments are held at different DC potentials during the sequence of ion capture and axial confinement. A DC potential applied to the first end-cap section (see Fig.~\ref{fig:UebersichtBeamline}) varies between 2 and 10\,V .
The second segment serves as a trapping region for the Sr$^{+}$ and Th$^{3+}$ ions. DC potentials in the range of $\pm$\,100\,mV are applied to the electrodes of this section in order to bring the ions to the RF minimum and to minimize the excess micromotion \cite{Berkeland:1998} for the Sr$^{+}$ ions. DC potentials applied to the last three sections also vary for different trapping regimes in the range of a few volts. 

The vacuum conditions in the trap decisively determine the lifetime of the highly reactive Th$^{3+}$ ions. The final stage of the differential pumping line reaches a base pressure in the range $\leq$\,10$^{-7 }$\,Pa. In addition, a liquid nitrogen cold trap is installed next to the ion trap to improve the quality of the vacuum. The materials for the entire vacuum system, especially around the ion trap, are selected for low magnetic permeability and low outgassing. We avoid the use of organic compounds and use titanium, 316L(N) stainless steel and ceramics like Al$_{2}$O$_{3}$ or ZrO$_{2}$.

The vacuum chamber provides two viewports along the trap axis and two viewports oriented at 45$^{\circ}$ from the trap axis for laser access. For the optical detection of the fluorescence signals a viewport on top of the vacuum chamber and a re-entrant viewport oriented perpendicularly to the trap are used.

\subsection{Optical setup}
The optical setup for the laser cooling and laser excitation of Th$^{3+}$ and Sr$^+$ ions is shown in Fig.\,2. The part related to Sr$^{+}$ consists of two external cavity diode lasers (ECDL) emitting at 461\,nm and 405\,nm for two-step photoionization of Sr \cite{Brownnutt:2007,Removille:2009}, a Sr$^{+}$ cooling laser at 422\,nm, and a distributed feedback laser (DFB) at 1092\,nm as a repumper (see, for example \cite{Berkeland:2002}). For the spectroscopy of Th$\,^{3+}$ two ECDLs at 690\,nm and 984\,nm are used (see Section IV). The radiation frequencies of the cooling and spectroscopy lasers can be stabilized via digital feedback from a scanning transfer cavity. An ECDL that is stabilized to a resonance of Rb in a vapor cell (see, for example \cite{Barwood:1991}) provides an optical reference for the cavity length stabilization and a calibration signal for the wavemeter (HighFinesse/\AA ngstrom WS7-60) used in the experiment to obtain absolute frequency measurements.

The two overlapped beams for the Sr ionization at 461\,nm and 405\,nm have a power of 5\,mW and 2\,mW, respectively, and a diameter of about 1\,mm in the trap center. The beams are directed at 45$^{\circ}$ angle to the second segment of the ion trap to be perpendicular to the Sr atomic beam emitted by the Sr oven. The Sr$^{+}$ cooling laser at 422\,nm is directed through the trap at an angle of a few degrees from the trap axis in order to cool both radial and axial motion. This cooling beam has approximately 0.5\,mm diameter and about 100\,µW power. The DFB repumper radiation is directed from the opposite side of the trap and is overlapped with the cooling beam at the trapping region. For the repumper a power of 4\,mW is used and its 
detuning is different from that of the cooling laser in order to avoid population trapping in a coherent superposition of Zeeman sub-levels that forms a dark state \cite{Removille:2009}.

\begin{figure}[h]
\includegraphics[scale=0.55]{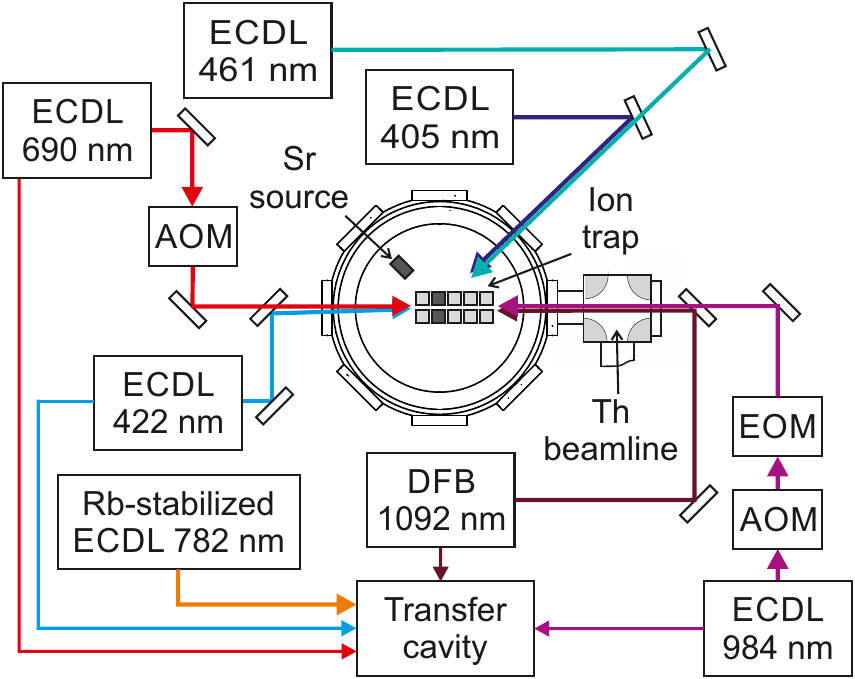}
\caption{\label{fig:optics} Optical setup for the laser spectroscopy of sympathetically cooled Th$^{3+}$ ions. Two external cavity diode lasers (ECDL) at 461\,nm and 405\,nm are used for two-step photoionization of Sr. Cooling of Sr$^{+}$ ions is provided by the ECDL at 422\,nm and the repumper at 1092\,nm. The ECDLs at 690\,nm and 984\,nm are used for the spectroscopy of Th$^{3+}$ ions. }
\end{figure}
 
The fluorescence of the S$_{1/2}$\,$\rightarrow$\,P$_{1/2}$ Sr$^{+}$ transition at 422\,nm is detected using an EMCCD camera and a photomultiplier tube (PMT). The EMCCD camera is used for the visualization of Sr$^{+}$ Coulomb crystals and to determine the number of stored Th$^{3+}$ ions (see section III). The PMT signal is used for the control of excess micromotion of the ion crystal and its readout is synchronized with the trap driving frequency. The PMT and the camera are equipped with a bandpass filter passing 422\,nm photons. 
For the observation of the Th$^{3+}$ ions, the optical path also contains a dichroic mirror at 45$^{\circ}$ incidence angle, which transmits the 422\,nm photons and reflects longer wavelength radiation. 
The Th$^{3+}$ fluorescence is detected using a PMT equipped with a 690\,nm bandpass filter. The readout of the PMT is gated by the 984\,nm laser pulse (see section IV).

\section{S\MakeLowercase{r}$^+$ -- T\MakeLowercase{h}$^{3+}$ two-species Coulomb crystals}
Direct laser cooling of Th$^{3+}$ has been demonstrated~\cite{Campbell:2011}, but the cooling power is limited because of the narrow ($<500$~kHz natural linewidth) cooling transitions. In addition, the hyperfine structure in the case of the $^{229}$Th isotope makes it necessary to apply multiple cooling laser sidebands in order to avoid population trapping in dark states. Therefore we investigate sympathetic laser cooling of Th$^{3+}$ with
$^{88}$Sr$^+$, as explained above. The charge-to-mass ratios of both species are similar, about 15\% higher for Th$^{3+}$, so that in a two-species Coulomb crystal the
thorium ions will occupy the region close to the axis of the RF ion trap, surrounded by one or several shells
of strontium ions. This configuration is ideal for reaching the whole ensemble of Th$^{3+}$ ions with a laser beam running along the trap axis, like for example a VUV beam for the excitation of the nuclear resonance \cite{Thielking:2023}. 
It is also convenient that because of their level structure, Sr$^+$ ions in their ground state will not interact resonantly with a VUV laser that drives the nuclear resonance. The ionization potential of Sr$^+$ of 11.03~eV is significantly higher than the $^{229}$Th isomer energy.

\begin{figure}
\includegraphics[scale=0.55]{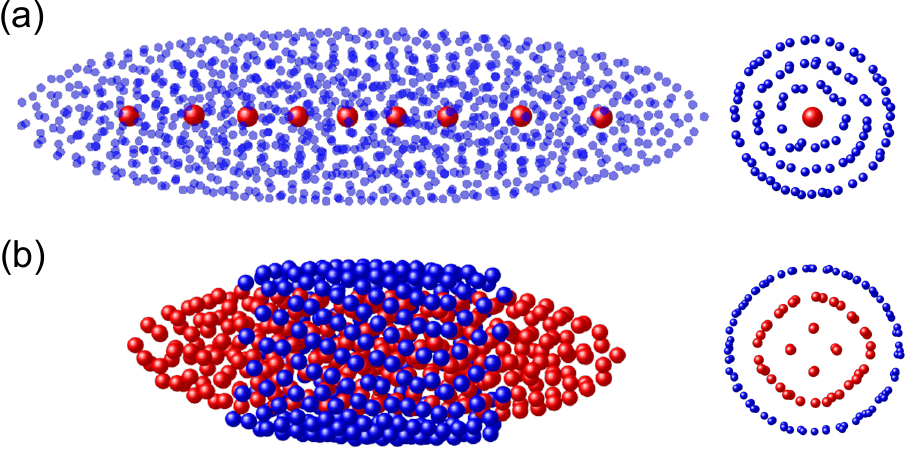}
\caption{\label{fig:Kristallsimu} Simulations of large two-species Coulomb crystals containing (a) 1000 Sr$^{+}$ (blue) and 9 Th$^{3+}$ ions (red), (b) 500 Sr$^{+}$ (blue) and 500 Th$^{3+}$ (red) ions. Images on the right are views along the axis of a slice of ions taken close to the center of the crystal.}
\end{figure}

Calculations of the resulting structures of two-species Coulomb crystals have been performed
for various ion numbers and trap parameters using the SIMION software ~\cite{SIMION:v8}. Typical structures
are shown in Fig.\,\ref{fig:Kristallsimu}, for a small number of 9 Th$^{3+}$ ions that arrange in a linear chain on the trap axis, and for a larger number of 500 Th$^{3+}$ ions that arrange in several shells, surrounded by a tubular arrangement of Sr$^+$. The simulations were performed with ion trap parameters similar to the experiments. They include the time dependence of the trap potential and the images shown are snapshots for a certain phase of the RF. Laser cooling is modeled by a velocity-dependent damping force that acts on both species. The low-temperature equilibrium structures that are shown here are determined by the Coulomb interaction and the trap potential alone and not influenced by the value of the damping parameters.

\begin{figure}
\includegraphics[scale=0.40]{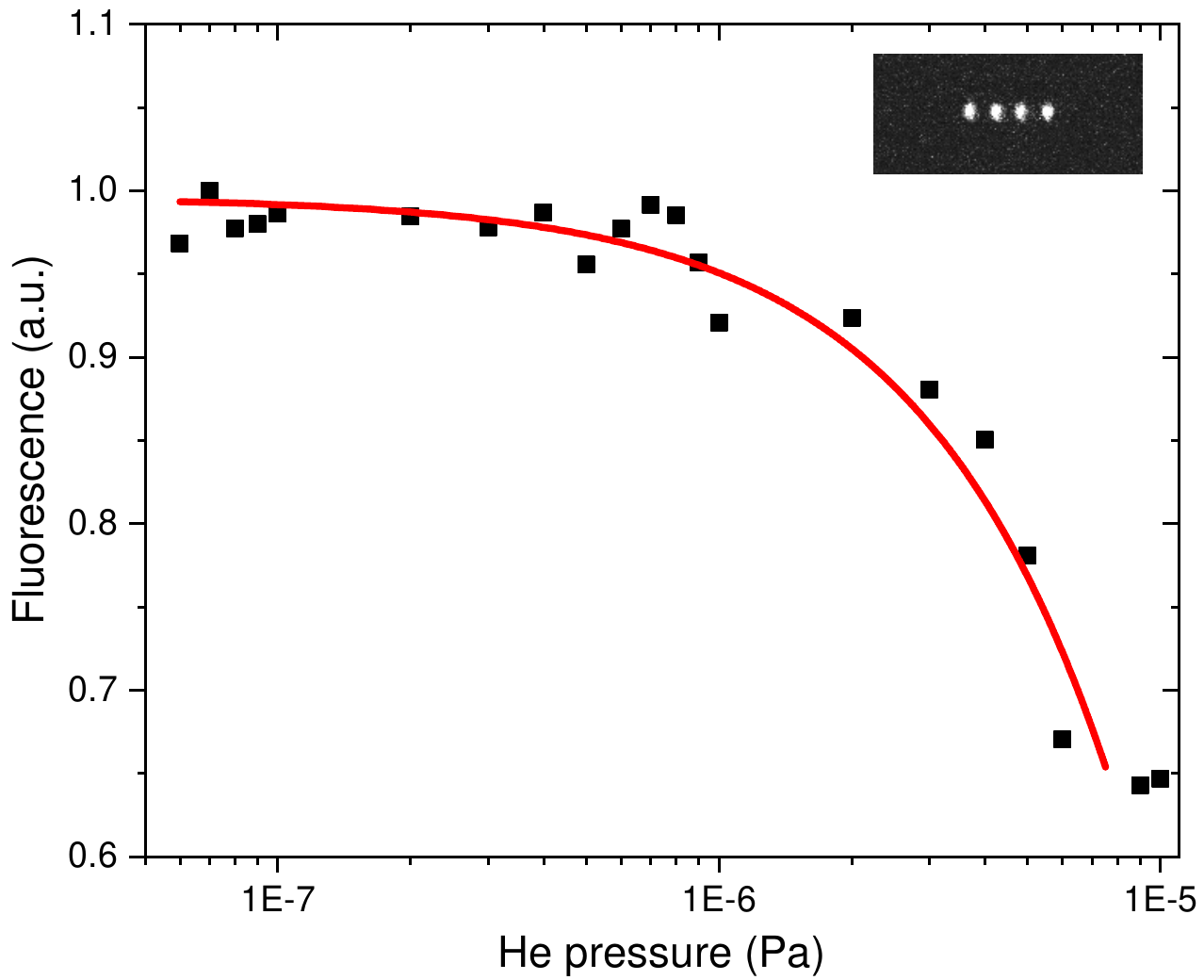}
\caption{\label{fig:DruckFluoresz} Fluorescence signal of Sr$^+$ ions as a function of helium buffer gas pressure. The solid line corresponds to a fit assuming the rate of collisions and heating in the Sr$^+$ Coulomb crystal to increase linearly in pressure. The two data points at the highest pressure deviate from the model: The Coulomb crystal has melted, leading to uncorrelated motion of the ions at a higher temperature. The inset shows an image of the Coulomb crystal.}
\end{figure}

\begin{figure}
\includegraphics[scale=0.65]{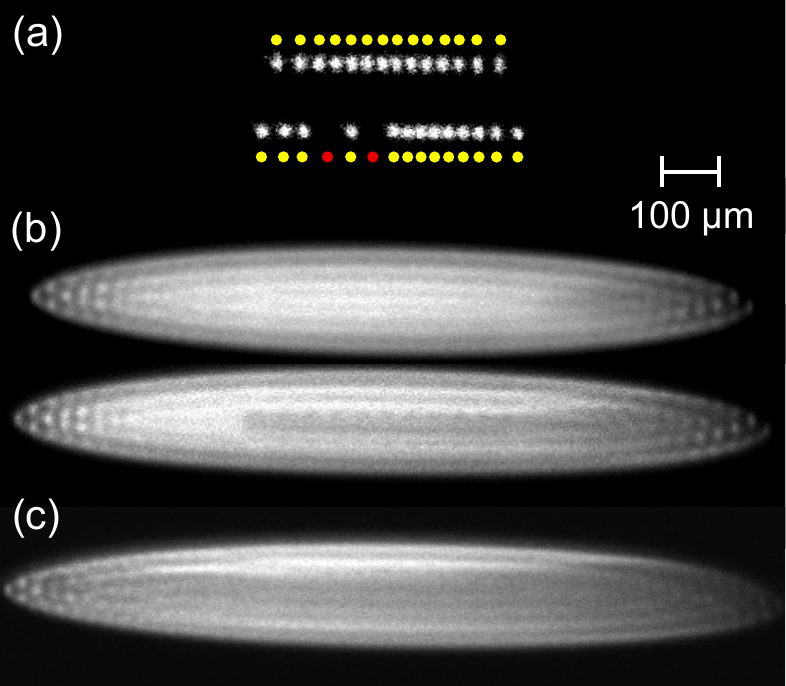}
\caption{\label{fig:experimental} Images of two-species Coulomb crystals observed on the 422\,nm Sr$^+$ fluorescence: (a) chain of Sr$^{+}$ ions before and after loading of two Th$^{3+}$ ions. The experimental EMCCD image is compared to simulated ion positions (colored dots, yellow: Sr$^+$, red: Th$^{3+}$), (b) Large Sr$^{+}$ Coulomb crystal before and after loading of about ten Th$^{3+}$ ions. The chain of dark regions on the axis indicates the positions of Th$^{3+}$. (c) Large Coulomb crystal of Sr$^+$ including more than 20 loaded thorium ions.}
\end{figure}

For the experiments on sympathetic laser cooling the ion trap is first loaded with a large $^{88}$Sr$^{+}$ cloud produced via two-step ionization of an atomic beam emitted by the Sr source. The amount of trapped ions is controlled by varying the dispenser oven temperature and the interaction time with the ionization lasers.
Depending on the trap parameters and the number of ions, one- or three-dimensional Coulomb crystals of Sr$^{+}$ can be observed.

As long as the gate valve between ion source and ion trap is open there is a residual helium pressure from the recoil ion source in the ion trap chamber. Therefore we investigated the influence of collisions with helium on the laser cooling of Sr$^+$.
A central elastic collision of a He atom with a cold Sr$^+$ ion will transfer 0.166 of the He kinetic energy, corresponding to about $1\times 10^{-21}$\,J at 300\,K. At a typical cooling laser detuning of 20\,MHz, it will take about $8\times 10^4$ scattered laser photons to extract this energy from the Sr$^+$ ion. With a laser at saturation intensity, this will take about 1\,ms. As long as the collision rate remains smaller than $10^{-5}$ of the photon scattering rate, the ion can be expected to be at the laser cooling temperature for most of the time. The Doppler limit for Sr$^+$ is 0.5\,mK. Figure 4 shows the strength of the fluorescence signal of a 4-ion linear chain of Sr$^+$ obtained from the EMCCD camera image as a function of He pressure. Under the assumption that each collision event temporarily reduces the fluorescence scattering rate until the Coulomb crystal is recooled, the dependence of the fluorescence signal of the ions on pressure $p$ is modeled as $I(p)/I(0)=1-\kappa p$. A fit with this function (solid line in Fig.\,\ref{fig:DruckFluoresz}, resulting in $1/\kappa$\,=\,$2.2\times10^{-5}$\,Pa) describes the experimental data well. 
At pressures above $8\times10^{-6}$\,Pa laser cooling was not sufficient to maintain a stable Coulomb crystal but the movement of the ions becomes uncorrelated (cloud phase \cite{Diedrich}). A similar behavior has been observed with laser-cooled Mg$^+$ ions where the transition from Coulomb crystal to cloud happened at a He pressure of $5\times10^{-6}$\,Pa \cite{Zhao:2006}.

After loading the Sr$^{+}$ Coulomb crystal a stopping voltage of +10\,V is applied to the electrodes of the first end-cap trap section for an efficient trapping of Th$^{3+}$ ions. A DC potential gradient of 0.2 V per trap segment is applied to the last three sections of the trap. The voltages applied to the trap segments 3 - 5 are 0.2\,V, 0.4\,V, and 0.6\,V respectively. Th$^{3+}$ ions are extracted as a $\sim$\,100\,µs bunch from the RFQ after $\approx$\,50\,s of accumulation. The voltage of the RFQ extraction lens is -5\,V. After the RFQ extraction pulse is applied, the blocking voltage of $+10$\,V of the ion trap focusing lens segment 6 (cf. Fig.\,\ref{fig:UebersichtBeamline}), is switched to $-5$\,V for 40\,µs with a delay of about 400\,µs. This protocol allows individual bunches of thorium ions to be injected into the trap and prevents reflection of the ions back to the ion beamline. 
During the loading, the frequency of the 422\,nm Sr$^+$ cooling laser is shifted to the red by a few hundred MHz from the optimal Doppler cooling frequency for a few seconds and then tuned back. This helps to improve the efficiency of trapping high-energy Th$^{3+}$ ions into a Sr$^{+}$ Coulomb crystal.  The computerized loading protocol allows us to add thorium ions to the crystal by repeating the injection sequence several times. The loading efficiency is estimated by detecting with a channeltron the amount of Th$^{3+}$ ions passing through the QMF: Typically 100 ions per bunch from the RFQ are detected here, while typically about 10 thorium ions are stored in a Sr$^{+}$ Coulomb crystal per bunch.

Figure\,\ref{fig:experimental} shows one- and three-dimensional Sr$^+$ Coulomb crystals before and after loading of thorium ions. Images of crystals with $^{229}$Th$^{3+}$ loaded from a $^{233}$U source and with $^{230}$Th$^{3+}$ loaded from $^{234}$U are similar. Since the ions are observed on the 422\,nm Sr$^+$ fluorescence, 
dark regions within the crystals indicate the positions of sympathetically cooled Th$^{3+}$ ions. The extension of these regions is larger than the distance between the neighboring Sr$^+$ ions because of the higher value of the positive charge. For the case of the linear ion chains, a numerical calculation of the equilibrium configuration in a static potential has been performed. The result (colored dots in Fig. 5) is in good agreement with the experimentally observed positions of the Sr$^+$ ions under the assumption that dark spaces in the chain are due to the presence of single ions of charge state $3+$.
The configuration in Fig.\,\ref{fig:experimental}(b) shows a linear chain of about 10 Th$^{3+}$ ions on the axis of a large Sr$^+$ Coulomb crystal. Loading of the 10 Th$^{3+}$ ions increases the length of the crystal by about 5\%. Fig.\,\ref{fig:experimental}(c) shows a different crystal containing more than 20 Th$^{3+}$ ions, likely in  a helical or tubular configuration. 
A loss of the Th$^{3+}$ ions from the crystals with a time constant of about 5000 s at the lowest pressure in the ion trap chamber is observed, significantly faster than for Sr$^+$ ions. This is in agreement with the observation that Th$^{3+}$ ions are highly reactive with molecules from the background gas, and that these reactions are accompanied with a charge exchange predominantly to molecular ions of charge state $2+$ \cite{Churchill}. The lower charge-to-mass ratio of these reaction products makes their ejection from the Coulomb crystal likely. 

\section{Laser spectroscopy of sympathetically cooled T\MakeLowercase{h}$^{3+}$ ions}
\begin{figure}[h]
\includegraphics[scale=0.85]{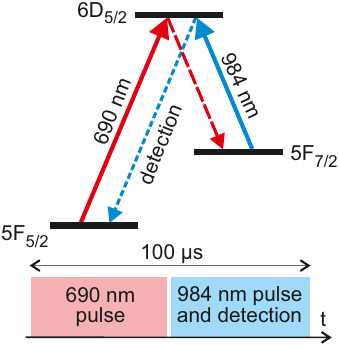}
\caption{\label{fig:detection} Scheme for pulsed excitation and fluorescence detection of Th$^{3+}$ ions. The 5F$_{7/2}$ metastable state is populated through the decay of the 6D$_{5/2}$ state by an excitation pulse at 690\,nm. The fluorescence signal at 690\,nm is registered during the 984\,nm repumper laser pulse in the absence of laser straylight at 690\,nm. The repetition rate for the excitation and detection pulse sequence is 10\,kHz.}
\end{figure}

For a demonstration of high-resolution laser spectroscopy of sympathetically cooled Th$^{3+}$ $\alpha$-recoil ions, we have chosen the isotope $^{230}$Th$^{3+}$ because it does not show a level splitting from hyperfine structure. We are probing the $\Lambda$-level scheme (see Fig. 6) that has already been studied with the isotopes $^{232}$Th$^{3+}$ and $^{229}$Th$^{3+}$ ~\cite{Campbell:2011}. 
Both transitions have relatively small oscillator strengths and spontaneous decay from the  $6D_{5/2}$ level leads  predominantly (branching ratio 7:1) to the emission of a 984~nm photon \cite{Biemont:2004,Safronova:2006}, while low-noise photon detectors with high quantum efficiency are more readily available for 690~nm. We have therefore implemented a pulsed excitation scheme that permits detection of the 690~nm fluorescence free from laser straylight. 
Both laser radiations at 690\,nm and 984\,nm are passed through acousto-optical modulators (AOM) for control of the temporal pulse shape. The laser beams are directed along the trap axis from opposite directions. The beam diameters are $\approx$\,0.5\,mm and the beams are overlapped with the Sr$^{+}$ lasers in the trapping region. The maximum pulse laser powers of the 690\,nm and 984\,nm ECDLs used in the experiment are in the range of a few hundred µW.
For the detection of the transition 5F$_{5/2}$\,$\rightarrow$\,6D$_{5/2}$ we apply a 30 - 50\,µs excitation pulse of the 690\,nm laser which populates the 5F$_{7/2}$ level through the decay of 6D$_{5/2}$ state (see Fig.\,\ref{fig:detection}). Next a pulse of the 984\,nm radiation is applied for 50 - 70\,µs. and the fluorescence signal at 690\,nm from the 6D$_{5/2}$ level decay to the ground state is registered during this period. The frequency of the 984\,nm laser can be modulated by a fiber-coupled electro-optical modulator (EOM) to provide a broad radiation spectrum. 

\begin{figure}[h]
\includegraphics[scale=0.37]{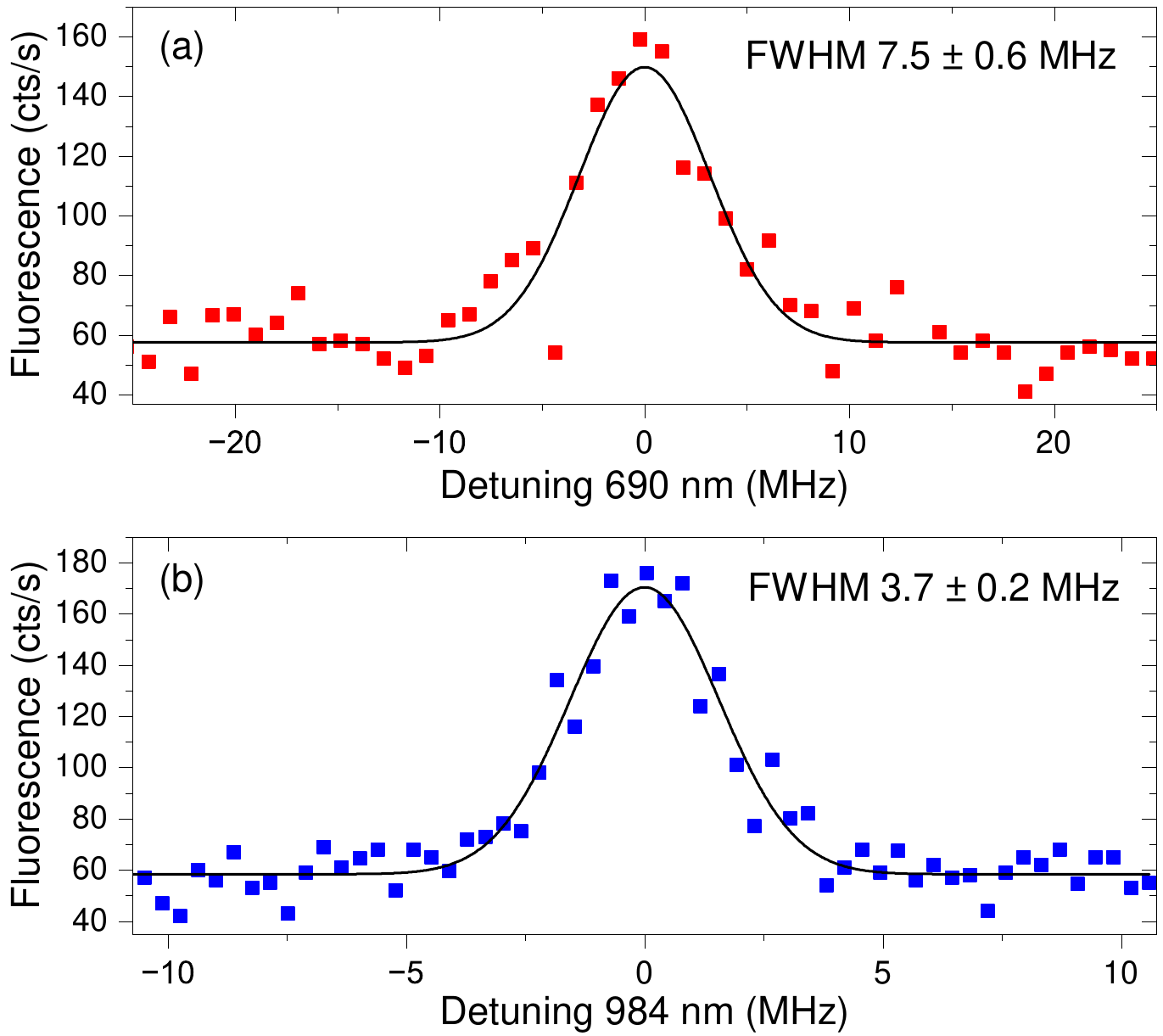}
\caption{\label{fig:Spektrum690984} Excitation spectra of sympathetically cooled $^{230}$Th$^{3+}$ ions: (a) the detected fluorescence signal is shown while scanning the 690\,nm laser frequency and keeping the 984\,nm laser frequency stabilized on the center of the 5F$_{7/2}$\,$\rightarrow$\,6D$_{5/2}$ transition via frequency lock to the transfer cavity, (b) shows the signal while scanning the frequency of the 984\,nm laser and keeping the 690\,nm laser frequency on the center of the 5F$_{5/2}$\,$\rightarrow$\,6D$_{5/2}$ transition. }
\end{figure}

A search for the frequencies of the $^{230}$Th$^{3+}$ lines is performed based on the known isotope shifts between $^{229}$Th and $^{232}$Th~\cite{Campbell:2011}. The search is provided by a scan of the 690\,nm laser in combination with a broadband EOM modulation of the 984\,nm laser.
After finding the center frequencies of both lines, the fluorescence decay signal at 690\,nm of the 5F$_{7/2}$\,$\rightarrow$\,6D$_{5/2}$ transition is detected by stabilizing one of the lasers at the center frequency and scanning the other laser over its transition. 
The recorded fluorescence signals of the two resonances are shown in Fig.~\ref{fig:Spektrum690984}. Coulomb crystals similar to the one shown in Fig.~5~(c) have been used, containing about 20 Th$^{3+}$ ions. The records are obtained with an integration time of 2\,s per data point. 
Fig.~\ref{fig:Spektrum690984}~(a) represents the fluorescence signal obtained by scanning the frequency of the 690\,nm laser while the frequency of the 984\,nm repumper is stabilized at the center of the 5F$_{7/2}$\,$\rightarrow$\,6D$_{5/2}$ transition. The resonance full width at half maximum (FWHM) shown in Fig.~\ref{fig:Spektrum690984}~(a) is about 7.5\,MHz. The excitation is provided by the 690\,nm and 984\,nm ECDLs pulse powers of 6\,µW and 80\,µW respectively. The trap drive voltage corresponds to the parameter $\textit{q}$ $\approx 0.2$.
As an alternative, we stabilize the frequency of the 690\,nm laser at the center of the 5F$_{5/2}$\,$\rightarrow$\,6D$_{5/2}$ absorption line and provide a scanning of the 984\,nm laser. The fluorescence signal observed while scanning the frequency of the 984\,nm laser is shown in Fig.~\ref{fig:Spektrum690984}~(b). The resonance has a FWHM of about 3.7\,MHz. The pulse power of the 984\,nm laser is about 50\,µW and the 690\,nm power is 30\,µW.

\begin{figure}[h]
\includegraphics[scale=0.35]{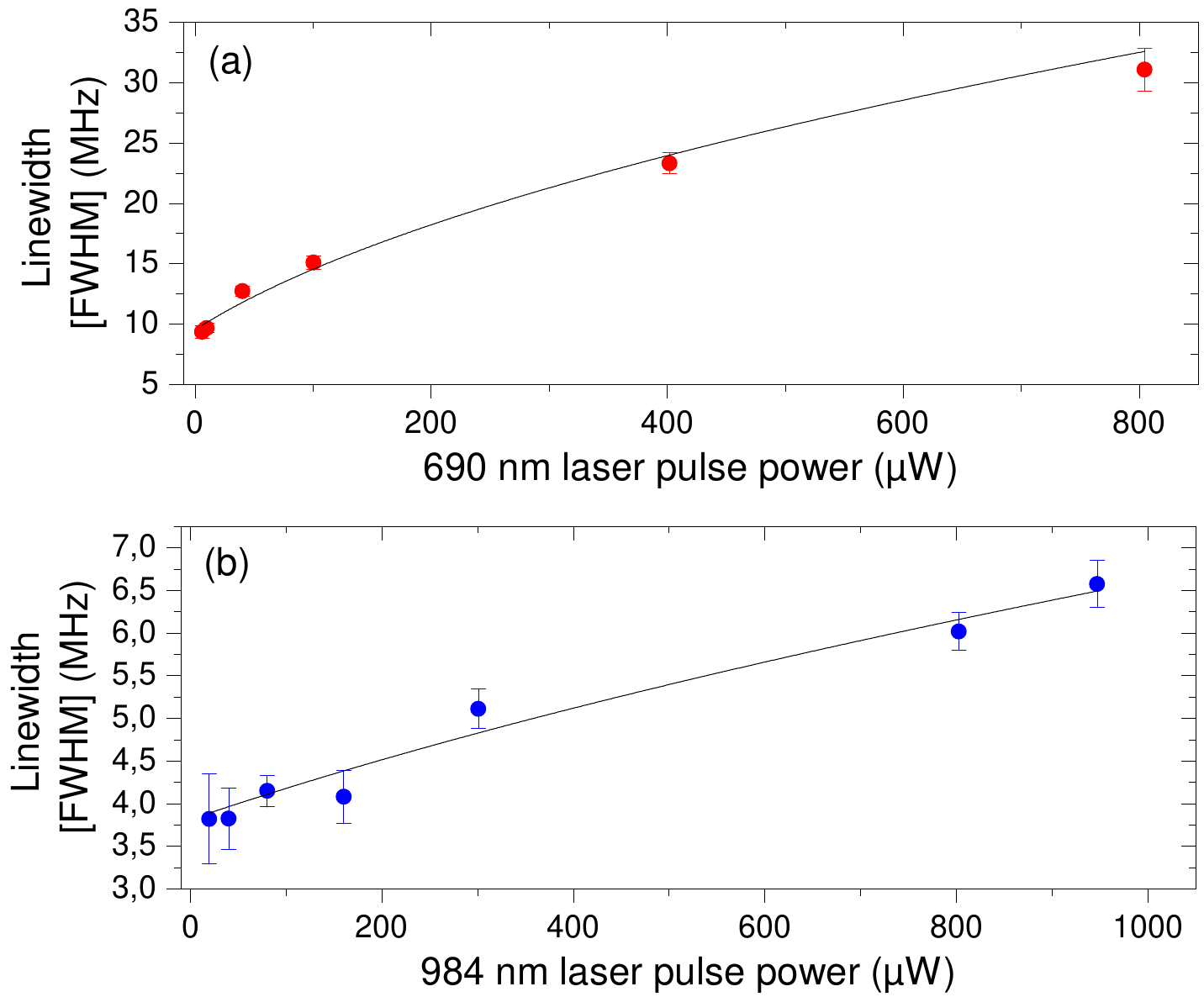}
\caption{\label{fig:PowerBroad} Saturation broadening of the fluorescence signal: (a) linewidth (FWHM) of the 690\,nm resonance vs. 690\,nm laser power while scanning its frequency and keeping the 984\,nm frequency stabilized on the center of its relevant transition. (b) shows similar measurements for the broadening of the resonance while scanning the 984\,nm laser frequency and keeping the 690\,nm laser frequency stabilized.}
\end{figure}

The lifetime of the 6D$_{5/2}$ state has been calculated to be in the range 0.3\,-\,0.7\,µs~\cite{Biemont:2004,Safronova:2006}, corresponding to a sub-MHz natural linewidth. Our estimations indicate that the spectroscopy lasers pulse powers in the range of $\sim$\,100\,µW can cause a significant saturation broadening of the excitation spectrum. Unresolved motional sidebands that are due to the driven micromotion of the Th$^{3+}$ ions in the Coulomb crystal can also contribute to the resonance broadening.
Therefore, we investigate saturation broadening effects and an influence of unresolved motional sidebands by changing of the laser powers and the trap drive voltage. Figure ~\ref{fig:PowerBroad} shows the dependence of the linewidth of the fluorescence signal on the pulsed laser power, for both the 690 nm and 984 nm lasers. The measurements are obtained by scanning one of the lasers over the resonance while the second laser frequency is stabilized on the center of the associated transition. For the experiment the excitation power of the scanning laser varies and the power of the frequency stabilized laser is kept low.
By changing the trap drive voltage in the $\textit{q}$ parameter range from 0.18 to 0.45, the width of the 690\,nm resonance slightly increases by a few MHz.

An estimation of the temperature of the sympathetically cooled thorium ions is provided using the linewidth of the observed fluorescence resonances at 690\,nm and 984\,nm with a minimized power broadening contribution shown in Fig.~\ref{fig:PowerBroad}. 
Tentatively assuming a thermal velocity distribution and the absence of other contributions to the linewidth apart from Doppler broadening, the wider of the two spectra in Fig.~\ref{fig:Spektrum690984} corresponds to a temperature of 140~mK, providing an upper limit for the temperature of the Th$^{3+}$ ions. 
This is above the temperature limit for Doppler cooling of Sr$^+$ of 0.5~mK, but still within the temperature range that is expected for sympathetically cooled Coulomb crystals in this type of ion trap \cite {Bowe:1999,Zhang:2007}. 

The measured absolute frequency of the resonance at 690\,nm is 434.28491(7)\,THz which corresponds to a wavelength of 690.31285(12)\,nm. The uncertainty is dominated by the uncertainty of the wavemeter and was checked in a comparison of two similar devices.  
Relative to the resonance frequency for $^{232}$Th$^{3+}$ \cite{Campbell:2009}, the isotope shift of the 5F$_{5/2}$\,$\rightarrow$\,6D$_{5/2}$ line for  $^{230}$Th$^{3+}$ is $-6.48(10)$\,GHz.
For the 5F$_{7/2}$\,$\rightarrow$\,6D$_{5/2}$ transition the absolute frequency of the  $^{230}$Th$^{3+}$ resonance is 304.61327(6)\,THz 
and the wavelength is 984.17400(21)\,nm. 
The negative isotope shift of this transition relative to $^{232}$Th is $-6.08(9)$\,GHz.
For both transitions the shift for the lighter isotope is negative (shifted to the red), 
as expected from field shift coefficients that have recently been calculated \cite{Dzuba:2023}. 
Using these coefficients and neglecting contributions from mass shift, the measured isotope shifts can be converted into a difference of the rms nuclear charge radius $\delta \langle r^{2} \rangle^{232,230}=0.187(3)$fm$^2$ between the isotopes. This result is in agreement with earlier data obtained from a transition in Th$^+$ ~\cite{Kaelber:1989}.

\section{Conclusion}
We describe in this paper an experimental apparatus for high-resolution spectroscopy of trapped Th$^{3+}$ recoil ions. We demonstrate the efficient loading of Th$^{3+}$ ions from uranium sources into a trap holding a large laser-cooled Sr$^{+}$ Coulomb crystal and the sympathetic cooling of the thorium ions into a two-species Coulomb crystal. The spectroscopic signal of the 5F$_{5/2}$\,$\rightarrow$\,6D$_{5/2}$ transition in $^{230}$Th$^{3+}$ is registered with a background-free detection method using pulsed excitation at 690\,nm. We have measured the absolute frequencies of the 5F$_{5/2}$\,$\rightarrow$\,6D$_{5/2}$ (690 nm) and 5F$_{7/2}$\,$\rightarrow$\,6D$_{5/2}$ (984 nm) transitions in $^{230}$Th$^{3+}$ and have calculated the isotopic shifts. The experimental apparatus will be used in the future for hyperfine spectroscopy of the ground and the isomeric state in $^{229}$Th$^{3+}$ ions, with the aim to obtain more precise values for the nuclear moments of these states \cite{Safronova:2013,Thielking:2018}. Finally, our demonstration of trapping Th$^{3+}$ at ultralow temperatures via sympathetic cooling enables these ions to serve as the reference of a  highly accurate nuclear optical clock \cite{Peik:2003}.

\section{Acknowledgements}
We would like to thank Thomas Leder and Martin Menzel for their expert technical support. We thank Lars von der Wense, Benedict Seiferle, and Peter G. Thirolf (LMU Munich) for the design of the buffer gas stopping cell and for providing the RF funnel. We also thank the LMU group for helpful discussions. We thank the Alpha and Gamma Spectrometry Group led by Stefan Röttger at PTB for providing an analysis of the uranium sources, and Tara Cubel Liebisch for a careful reading of the manuscript. 
This work has been funded by the European Research Council (ERC) under the European Union’s Horizon 2020 research and innovation programme (Grant Agreement No. 856415), the Deutsche Forschungsgemeinschaft (DFG) – SFB 1227 - Project-ID 274200144 (Project B04), and by the Max-Planck-RIKEN-PTB-Center for Time, Constants and Fundamental Symmetries.

\end{document}